\documentclass[12pt,onecolumn]{IEEEtran}
\usepackage[colorlinks,bookmarksopen,bookmarksnumbered,citecolor=red,urlcolor=red,]{hyperref}
 \usepackage{CJK}
 \usepackage{indentfirst}
\usepackage{algorithm,algorithmic,amsbsy,amsmath,amssymb,epsfig,bbm,mathrsfs, bbm} 
\usepackage{multirow}
\usepackage{mathrsfs}
 \usepackage{epstopdf}

\usepackage{verbatim}
\usepackage{amsfonts}
\usepackage{amsthm}
\usepackage[subfigure]{graphfig}
\renewcommand{\baselinestretch}{1.6}

\begin{document}

\title{Adaptive Power Management for Wireless Base Station in Smart Grid Environment}

\author{Dusit Niyato, Xiao Lu, and Ping Wang \\
School of Computer Engineering, Nanyang Technological University, Singapore.
\thanks{\textbf{D. Niyato} is the corresponding author (Email: dniyato@ntu.edu.sg).}}

\maketitle

\begin{abstract}
The growing concerns of a global environmental change raises a revolution on the way of utilizing energy. In wireless industry, green wireless communications has recently gained increasing attention and is expected to play a major role in reduction of electrical power consumption. In particular, actions to promote energy saving of wireless communications with regard to environmental protection are becoming imperative. To this purpose, we study a green communication system model where wireless base station is provisioned with a combination of renewable power source and electrical grid to minimize power consumption as well as meeting the users' demand. More specifically, we focus on an adaptive power management for wireless base station to minimize power consumption under various uncertainties including renewable power generation, power price, and wireless traffic load. We believe that demand-side power management solution based on the studied communication architecture is a major step towards green wireless communications.
\end{abstract}

\section{Introduction}

Until now, wireless communication system has been well developed and optimized in terms of spectrum efficiency, transmission reliability, and users' satisfaction from a variety of mobile applications. However, the new challenge of wireless communication system recently emerges due to the increasing power cost and higher volume of teletraffic demand. These create an immediate need for the `green' wireless communications which is a set of concepts, designs, and approaches to improve power efficiency of wireless system, while meeting the quality-of-service (QoS) of mobile users. The green wireless communications will help the network operator not only to save the power cost through better power efficiency per service, but also to provide the environmental responsibility through minimizing the environmental impact (e.g., by using renewable power source to reduce the ${\mathrm{CO}}_2$ emission). In addition, wireless communication system has to adapt to the change of power supply side which will become more dynamic and distributed, known as `smart grid'. Given all these requirements, the issue of power management for wireless system will become crucial and needs to be addressed accordingly.

The typical wireless communication system consists of three parts, i.e., core network, access network, and mobile unit. The largest fraction of power consumption in wireless networks comes from access network, especially the wireless base station, as the number of wireless base stations is enormous and the corresponding power consumption is high. Statistics show that the wireless base stations are responsible for more than half of the power consumption in wireless communication system worldwide. In contrast, the total power consumption of network cores is low due to the small quantity of core devices, and mobile terminals consume even less power because of their mobile nature. With this premises, power saving in wireless base station is particularly important for network operator.

In this article, we first provide an introduction of green wireless communications with the focus on the power efficiency of wireless base station, renewable power source, and smart grid. Then, we consider an adaptive power management for wireless base station with a renewable power source in smart grid environment. While the main power supply of wireless base station is from electrical grid, a solar panel is considered to be an alternative power source. An adaptive power management is used to coordinate among electrical grid and solar panel, which is energy-efficient and allows for greater penetration of variable renewable energy sources in green communication system. With smart grid, the adaptive power management can communicate about the power price with electrical grid and adjust power buying accordingly. However, in such an environment, many parameters are uncertain (e.g., generated renewable power, power price from electrical grid, and power consumption of wireless base station which depends on the traffic load). Therefore, the stochastic optimization problem is formulated and solved to achieve the optimal decision of power management. The performance evaluation is performed and clearly shows that with the optimal policy of adaptive power management, the power cost of the wireless base station can be minimized.

The rest of this article is organized as follows. Section~\ref{sec:overview} presents an overview of green wireless communications. Section~\ref{sec:adaptive} introduces the adaptive power management for wireless base station with renewable power source in smart grid environment. Finally, Section~\ref{sec:conclusion} concludes the article.


\section{Green Wireless Communications}
\label{sec:overview}

Lots of research and development efforts have been made in wireless industry, aiming for environment-friendly power solutions which lead to green wireless communications. Green wireless communications will contribute to the reduction of our global carbon footprint and enable mutual broader impacts across related fields, among which, renewable power resources and smart grid are attracting growing interests. In this section, we present an overview of the major concerns in green wireless communications related to renewable power and smart grid. 

\subsection{Renewable Power Sources and Green Wireless Communications}

The climate change deriving from enormous power consumption as a result of rapid industrial development has pushed people to use sustainable alternative power, among which, renewable power resources with their low pollution and sustainable accessibility are attractive as a placement of traditional fossil energy. In green wireless communications, renewable power sources can be used to replenish the energy of wireless base station/network nodes as alternatives to traditional power source. Manufacturers and network operators have started developing and deploying wireless base station with renewable power source~\cite{Gozalvez2010}. For example, Ericsson and Telecom Italia developed and tested the Eco-Smart solution which uses solar panel to fully power the cell site. Vodafone, China Mobile and Huawei jointly performed various experiments on the renewable power sources including solar panel, wind power generator, and the hybrid system for wireless base station~\cite{belfqih2009}. The experiments focus on the implementation verification, power reliability, and cost reduction.

However, renewable power sources, e.g., solar, wind, hydro, geothermal, tidal energy and biomass, are typically featured as weather-driven, uneven geographically distributed, non-scheduled, and relatively unpredictable. Thus, the main problem of the applications of renewable energy is that the power generation cannot be fully forecasted and may not follow the trend of actual power demand. 

For the use of renewable power source in green wireless communications, efficient power management is the primary concern since replenishment rate depends on the renewable power generation which is known to be fluctuant and intermittent. In wireless network, power saving often requires a degradation in network performance (i.e., higher latency and lower throughput). Designing efficient power management is therefore challenging in compromising between power saving and network performance. A number of interesting works have been carried out to address the issue. \cite{Zhang05} studied throughput and value-based wireless transmission scheduling under time limits and energy constraints for wireless network employing a renewable power source. Optimal scheduling algorithms that selectively transmit data at calculated rates were presented to maximize the throughput and total transmission value. \cite{Lin07} developed a model to characterize the performance of multi-hop wireless network with different type of energy constraints (i.e., renewable and non-renewable power source). Based on this model, the authors proposed an algorithm of energy-aware routing with distributed energy replenishment to optimally utilize available energy. More recently, \cite{Badawy10} proposed an energy-aware resource provisioning algorithm for energy provision in solar-powered wireless mesh network with the aim to save resource while preventing node outage. Significant resource saving is achieved with the proposed algorithm in hybrid networks with a mixture of solar-powered and continuously powered nodes.

\subsection{Smart Grid and Green Wireless Communications}

Large-scale centralized electricity generation and high-voltage long-distance transmission adopted by traditional electrical grid system are the two basic causes of power inefficiency. The concept of smart grid is introduced by using information and communications technology (ICT) to improve the efficiency and reliability of the electrical grid. The main features of smart grid related to the green wireless communications are the demand side management (DSM), decentralized power generation, and price signaling. With demand side management, the power generators and consumers can be interacted to improve the efficiency of power supply and consumption. For example, the operation and power consumption of the deferrable load (e.g., heating and pumping) can be adjusted according to the generator capability. With decentralized power generation, the power generation can be performed by consumers and small power plant (e.g., solar panel and wind turbine). As a result, consumers will be less dependent to the main electrical grid, reducing the power cost and avoiding impact from power failure. With price signaling, the consumers will be aware of the current power price and the generator can use cheap power price to encourage the consumers to use the electric power during off-peak period (e.g,. night time or weekend). Consequently, the peak load will be reduced which results in lower investment for the infrastructure (e.g., transmission line and substation).

Many current research works focus on the enabling technologies of interaction between smart grid and wireless communications~\cite{Heile10}. On one hand, wireless communications is a key component in smart grid to communicate a variety of data and measurement among power generators, transmission lines, distribution substations, and consumer loads. On the other hand, smart grid can be used to support green wireless communications for the better use of power to provide wireless service to mobile users. This similar concept has been explored in the ``green computing''~\cite{MohsenianRad2010}. In this case, the data center can schedule the service request (i.e., data processing) according to the power supply from electrical grid. Also, efforts have been made on theoretical analysis. In wireless network, each wireless base station/node powered by smart grid might be selfish in improving performance in capacity or QoS. How to improve power saving without adversely affecting QoS and capacity is one of the main concerns. Recent progress in wired distributed computing theory~\cite{Jai05} provided fundamental models for coordinated management and load balancing of wireless base stations underneath smart grid, which has potential for addressing the concern.

From above related works, it is clear that the use of renewable power source and smart grid will be the major trend of the green wireless communications. However, there are many issues to be addressed including the protocol design, radio resource optimization, and power management.


\section{Adaptive Power Management}
\label{sec:adaptive}

In this section, the adaptive power management for wireless base station with renewable power source in smart grid environment is presented. First, the system model is described. The optimization formulation to achieve optimal policy of adaptive power management is discussed. Then, the performance evaluation of the proposed scheme is presented.

\subsection{System Model}

\begin{figure}[h]
\begin{center}
\includegraphics[width=3.75in]{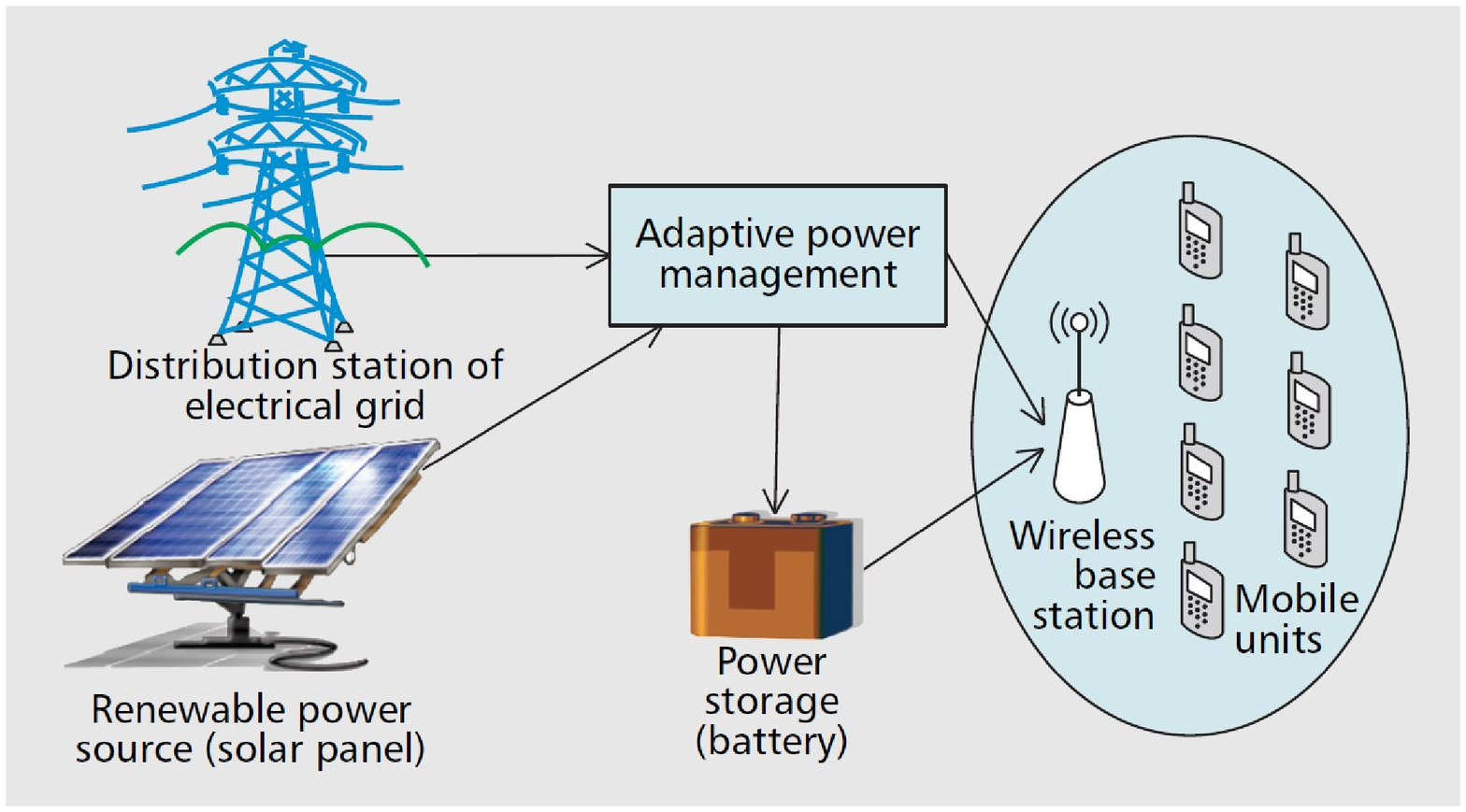}
\caption{System model of adaptive power management for wireless base station in smart grid.}
\label{fig:systemmodel}
\end{center}
\end{figure}

The system model of adaptive power management for wireless base station is shown in Fig.~\ref{fig:systemmodel}. The components in this system model are as follows:
\begin{itemize}
	\item {\em Wireless base station:} Wireless base station or access point is a centralized device used to provide wireless services to mobile units. Wireless base station is the power consumption device. The amount of power consumption depends on the type of base station and traffic load (i.e., the number of ongoing connections from active mobile units).
	\item {\em Electrical grid:} Electrical grid provides an interconnected network including transmission lines and distribution substations for delivering electricity from generators to consumers. Electrical grid is a main source of power supply to the wireless base station. The power supplied from electrical grid has a price per kWh (kilowatt-hours).
	\item {\em Renewable power source:} Renewable power is provided from the natural resources such as sunlight and wind which are replenishable. As a result, the variable cost of renewable power is cheaper than that from electrical grid. However, the power generated by renewable source is typically random due to the unpredictable availability of natural resources. Therefore, renewable source is considered to be an alternative power supply of electrical grid. The maximum amount of generated power from renewable source (i.e,. capacity) is denoted by $R$ kW (kilowatt).
	\item {\em Power storage:} Battery is the power storage device for wireless base station. Battery can be charged by the power from the renewable power source or from the electrical grid when the power price is low. Battery has a limited maximum capacity for power storage denoted by $B$ kW. Note that the power stored in a battery can decrease even without consumption. This is referred to as the self-discharge phenomenon. The self-discharge rate per time unit is denoted by $L$.
	\item {\em Adaptive power management controller:} Adaptive power management controller has a mechanism to make decision on power supply from renewable source and electrical grid to the battery and wireless base station. Adaptive power management controller utilizes the available information to optimize the decision with the objective to minimize power cost while meeting the demand of wireless base station. The details of this optimization will be presented later in this section.
\end{itemize}

While the electrical grid is owned by the utility company, wireless base station, power storage, renewable power source, and adaptive power management controller belong to the network operator with the objective to minimize power cost. The information to be exchanged and maintained among above components to support adaptive power management of wireless base station is as follows: price of power from electrical grid, generated power from renewable source, battery storage, and power consumption of wireless base station. These information is measured and reported periodically to the adaptive power management controller. In the smart grid environment, the communications infrastructure to transfer these information is assumed to be available. In this case, broadband access (e.g., ADSL) and local area network (e.g., Ethernet) can connect the adaptive power management controller with the electrical power grid and renewable power source as well as wireless base station.

Adaptive power management can be considered as the demand side management (DSM). DSM is part of smart grid which allows the power consumers (e.g., wireless base station) to adjust their power consumption. The main aim of DSM is to balance the consumption from peak to non-peak period such that the cost of infrastructure to accommodate peak demand is reduced. In this context, adaptive power management has ability to control the power buying from electrical grid (i.e., consumption) given the varied price and power generated from renewable source. In addition, adaptive power management will defer power buying when the price is high (e.g., peak hour) and move the consumption (e.g., charging battery) to the off-peak duration such as nighttime. 

The power consumption models of a wireless base station have been studied in the literature, e.g.,~\cite{arnold2010,richter2010}. For wireless base station, the power consumption is composed of two parts, i.e., static and dynamic. Static power consumption is constant when the base station is active even without active connections from users. On the other hand, dynamic power consumption depends on the active connections and is a function of traffic load. The power consumption of base station is from different components including power amplifier, signal processing unit, antenna, and cooling. In this article, the micro base station is considered in which its power consumption depends on the traffic load (i.e., dynamic power consumption is significant compared to static consumption). The power consumption of the micro base station can be expressed as follows: $C = E_{\mathrm{st}} + E_{\mathrm{dy}} N$ where $E_{\mathrm{st}}$ is the static power consumption, $E_{\mathrm{dy}}$ is the dynamic power consumption coefficient, and $N$ is the number of active connections. The static and dynamic power consumption of the micro base station depends on the transmit power, power amplifier efficiency, and power supply loss. Also, the dynamic power consumption depends on the signal processing and transmit power per connection. In summary, given the number of active connections, the power consumption of wireless (i.e., micro) base station can be calculated. This information will be used to optimize the decision by the adaptive power management controller.

\subsection{Optimization of Adaptive Power Management}

Adaptive power management of wireless base station is a challenging issue due to the uncertainty in the environment and system. To address this issue, the stochastic optimization problem can be formulated and solved to obtain the best decision of the adaptive power management controller such that the power cost of wireless base station is minimized.

\subsubsection{Uncertainty}

A variety of uncertainties exist for the power management for wireless base station.
\begin{itemize}
	\item {\em Renewable power source:} The power generated from renewable sources such as solar and wind generators is highly random due to the weather condition~\cite{Ponnambalam2010}. For example, the solar energy depends on the amount of sunlight. Cloud and rain which are unpredictable weather reduce the amount of generated power.
	\item {\em Power price from electrical grid:} Due to the unpredictable condition (e.g., demand) of the electrical power grid, the power price can be random within a certain range depending on the current system conditions~\cite{heredia2010}. For example, power price can be high (i.e., peak-hour price) in a certain time period. In this case, consumer can be informed with power price (i.e., price signaling feature in smart grid).
	\item {\em Traffic load of wireless base station:} The connection arrival (i.e., newly initiated and handoff users) can be varied (e.g., due to the mobility). Also, the connection demand of wireless base station depends on the usage condition (e.g., special event which results in peak load). As a result, the number of ongoing connections $N$ will be random~\cite{Epstein2000}, and the power consumption which can be obtained from the aforementioned power consumption model is also random.
\end{itemize}

The uncertainty can be represented by the ``scenario'' which is the realization of random variable. The scenario takes value from the corresponding space which is commonly assumed to be finite discrete set. For example, the power price at a certain period can take value from a set of 12 and 20 cents per kWh (i.e., normal and peak-hour prices, respectively). The scenario can be also defined over multiple periods. For example, with three periods in one day, the first scenario can be defined as $\{ 12, 12, 20, 12 \}$ cents per kWh for the power prices in the morning (6:00-12:00), afternoon (12:00-18:00), evening (18:00-24:00), and at night (0:00-6:00), respectively. Alternatively, the second scenario can be defined as $\{ 12, 20, 20, 12 \}$ cents per kWh. That is, the second scenario represents the case of having peak-hour price in the afternoon. With multiple random parameters, the scenario is defined as a composite value of generated renewable power, power price from electrical grid, and power consumption of wireless base station. For example, one scenario denoted by $\omega$ is defined as follows: For morning, afternoon, evening, and night, the generated renewable powers are $\{ 130, 290, 0, 0\}$ W, the power prices are $\{ 12, 12, 20, 12\}$ cents per kWh, and power consumption are $\{ 200, 230, 240, 200\}$ W, respectively. The scenarios can be extracted from historical data, e.g., the traffic load history and power price from electrical grid. Also, the weather forecast can be used to determine the scenario of generated renewable power. 

The probability distribution associated with the scenarios of generated renewable power, power price from electrical grid, and power consumption of wireless base station can be estimated. Given the observation period (e.g., 60 days), the number of days for the observed scenario can be counted. The corresponding probability can be then calculated by dividing this number of days by the duration of observation period (i.e., 60 days). For example, if the number of days for the power price scenario $\{ 12, 12, 20, 12\}$ is 15 days, while the number of days for scenario $\{ 12, 20, 20, 12 \}$ is 45 days, then the probabilities for the first and second power price scenarios are $15/60 = 0.25$ and $45/60 = 0.75$, respectively. The same method can be applied for the scenarios of generated renewable power and power consumption.

Given the uncertainty, the objective of adaptive power management controller is to minimize the power cost buying from the electrical grid with the constraint to meet the power consumption demand of wireless base station. 

\subsubsection{Stochastic Programming Formulation}

To obtain the decision of adaptive power management controller under uncertainty, the optimization problem based on multi-period linear stochastic programming can be formulated and solved~\cite{birge1997}. Stochastic programming is a mathematical tool to model the optimization problem with uncertainty of parameters. Stochastic programming is an extension of the deterministic mathematical programming in which the stochastic programming does not have a strong assumption on the complete knowledge of the parameters. Instead, for stochastic programming, the probability distribution of random parameters which can be estimated is incorporated into the optimization formulation. Stochastic programming can be used to obtain the optimal solution which is a feasible policy for the possible cases (i.e., scenarios). This optimal solution or policy which is a mapping from scenario to the decision will minimize the expectation of the objective (i.e., cost). The optimal solution of stochastic programming can be obtained by formulating equivalent deterministic mathematical program in which the standard methods (e.g., interior point method) can be applied efficiently. For example, linear stochastic programming problem can be transformed into the deterministic linear programming problem and solved to obtain the optimal solution.

Although there exist other approaches (e.g., Markov decision process, robust optimization, and chance-constrained programming) to address the optimization problem with uncertainty, these approaches are not suitable for the cost optimization of adaptive power management for wireless base station. For Markov decision process, the stochastic process of the random parameters must have Markov property. That is, the the state (i.e., scenario) of the random parameter depends on the current state, but not past state. This Markov property may not be held in many situations (e.g., power price of electrical grid). For robust optimization, the solution is obtained only for the worst case scenario with which performance can be unrealistically poor due to the consideration of extreme case. For chance-constrained programming, with optimal solution, the constraint violation will be bounded by the threshold. However, only complex analysis exists for the basic probability distribution (e.g., normal distribution). 

Therefore, stochastic programming becomes the best approach for the adaptive power management since this approach can be used to obtain the optimal solution which ensures that all constraints will be met. The efficient method can be applied to obtain the optimal solution for possible scenarios in which the expectation of cost given uncertainty is minimized.

The multi-period stochastic programming model for adaptive power management is shown in Fig.~\ref{fig:optimization}. This optimization model is for a decision horizon which is divided into $T$ decision periods. We consider the length of a period to be one hour in which the spot power price from electrical grid can be varied. The objective and constraints of optimization formulation defined in Fig.~\ref{fig:optimization} are defined as follows:
\begin{itemize}
	\item (\ref{eq:objective}) is an objective to minimize the expected cost due to power buying from electrical grid and battery loss due to self discharging over the entire decision horizon (i.e., $t=1,\ldots,T$), where ${\mathbb{E}} (\cdot)$ is expectation, and $P_{t,\omega}$ is a power price. This expectation is over all scenarios in space $\Omega$ given the corresponding probability $Pr(\omega)$ of scenario $\omega \in \Omega$. 
	\item (\ref{eq:const1}) is a constraint for the balance of power input and output of a decision period $t$. The power input of a decision period $t$ includes the power stored in battery $s_{t,\omega}$ at the beginning of period $t$, power buying from electrical grid $x_{t,\omega}$ and generated renewable power $R_{t,\omega}$ in period $t$. The power output of a decision period $t$ includes the power remained in battery $s_{t+1,\omega}$ at the end of period $t$ (i.e., at the beginning of period $t+1$), the power consumption of wireless base station in the current period $C_{t,\omega}$, and excess power $y_{t,\omega}$. Note that the excess power is used to represent the amount of power input exceeding the power consumption and battery capacity. 
	\item (\ref{eq:const2}) is a constraint of power storage that the power in the battery must be lower than or equal to the capacity $B$.
	\item (\ref{eq:const3}) is the initial and termination condition constraint where $B_1$ and $B_T$ are the power storage in battery at the first and last decision periods, respectively.
	\item (\ref{eq:const4}) is the constraints of non-negative value of power.
\end{itemize}

The multi-period stochastic programming model can be transformed in to the linear programming problem~\cite{birge1997}, and the standard method of solving linear programming can be applied to obtain the solution. The solution is the amount of power buying from electrical grid denoted by $x^*_{t,\omega}$ at time period $t$ given scenario $\omega$. This solution is applied when the realization of the scenario of generated renewable power, power price, and power consumption is observed. 

\subsection{Performance Evaluation}

\subsubsection{Parameter Setting}

We consider an adaptive power management for a micro base station. The parameter setting of the micro base station is similar to that in~\cite{arnold2010}. The static power consumption is $E_{\mathrm{st}} = 194.25W$, while the dynamic power consumption coefficient is $E_{\mathrm{dy}} = 24W$ per connection. The transmission range of micro base station is 100 meters in which the transmit power calculated as in~\cite{arnold2010} is applied to ensure the reliable connectivity of the users. The maximum number of connections of the micro base station is 25. 

We consider the solar panel as a renewable power source. The capacity of the solar panel is 300Wh. The battery capacity is 2kW. The initial and termination power levels of battery are assumed to be 500W. The self-discharge rate of a battery is 0.1\% per hour. We consider the randomness of the power price, generated renewable power, and the traffic load of the micro base station. For power price, two scenarios are considered, i.e., peak-hour and normal prices, whose average power prices are 20 and 12 cents per kWh, and the corresponding probabilities are 0.6 and 0.4, respectively. For renewable power, two scenarios are considered, i.e., clear sky and cloudy, whose average generated power from 6:00-18:00 are 195Wh and 100Wh, and the corresponding probabilities are 0.6 and 0.4, respectively. For traffic load of the micro base station, five scenarios are considered, i.e., heavy uniform, medium uniform, light uniform, heavy morning, and heavy evening, and the corresponding probabilities are 0.1, 0.1, 0.2, 0.2, and 0.4, respectively. For heavy uniform, medium uniform, and light uniform scenarios, the traffic load is uniform and the mean connection arrival rates are 0.56, 0.22, and 0.15 connections per minute, respectively. For heavy morning and heavy evening scenarios, the connection arrival is peak during 8:00-11:00 and 17:00-21:00, whose mean connection arrival rates is 0.8 connections per minute, respectively. The adaptive power management scheme is optimized for 24 hour period.

\subsubsection{Numerical Results}

\begin{figure}[h]
\begin{center}
\includegraphics[width=4.3in]{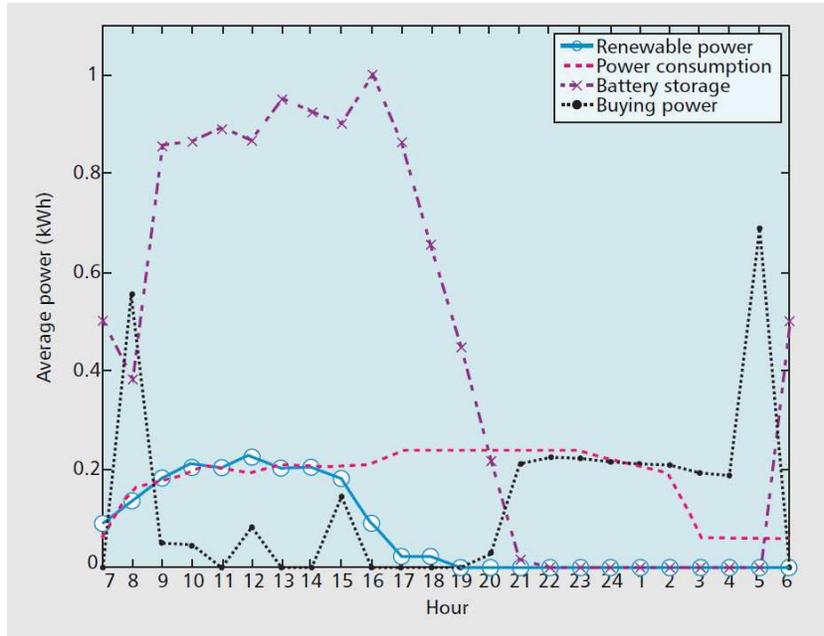}
\caption{Average generated renewable power, power consumed by wireless base station, battery storage, and buying power from smart grid.}
\label{fig:result_power_time}
\end{center}
\end{figure}

Fig.~\ref{fig:result_power_time} shows the different average power over the optimization period. In this case, the renewable source (i.e., solar panel) can generate power only when the sunlight is available. Therefore, the adaptive power management controller has to optimize the power storage in the battery and the power buying from electrical grid to meet the requirement of the micro base station. We observe that the battery is charged with the renewable power. The power is bought from electrical grid occasionally for the micro base station (e.g., when the renewable power is not available) or to charge the battery (e.g., at 8:00). Given the average power shown in Fig.~\ref{fig:result_power_time}, the power cost of this micro base station with adaptive power management is 12.79 dollars per month.

For comparison purpose, we consider a simple power management scheme in which the power is bought from electrical grid when there is no renewable power. In this case, the power storage in battery is maintained to be constant (i.e., 1kW). The power cost per month of this simple power management scheme is 14.7 dollars per month. Clearly, the proposed adaptive power management scheme achieves lower cost, and, hence, lower power consumption from electrical grid. Nevertheless, even though the cost saving for one micro base station may be marginal (i.e., 1.91 dollars per month or about 13\%). However, this cost saving can be significant when a number of micro base stations are deployed (e.g., 100 base stations in a campus). Also, the less use of power from electrical grid which is mostly generated from the fossil fuel (e.g., coal and oil) will reduce the ${\mathrm{CO}}_2$ emission which is the main aim of green wireless communications.

\begin{figure}[h]
\begin{center}
\includegraphics[width=4.3in]{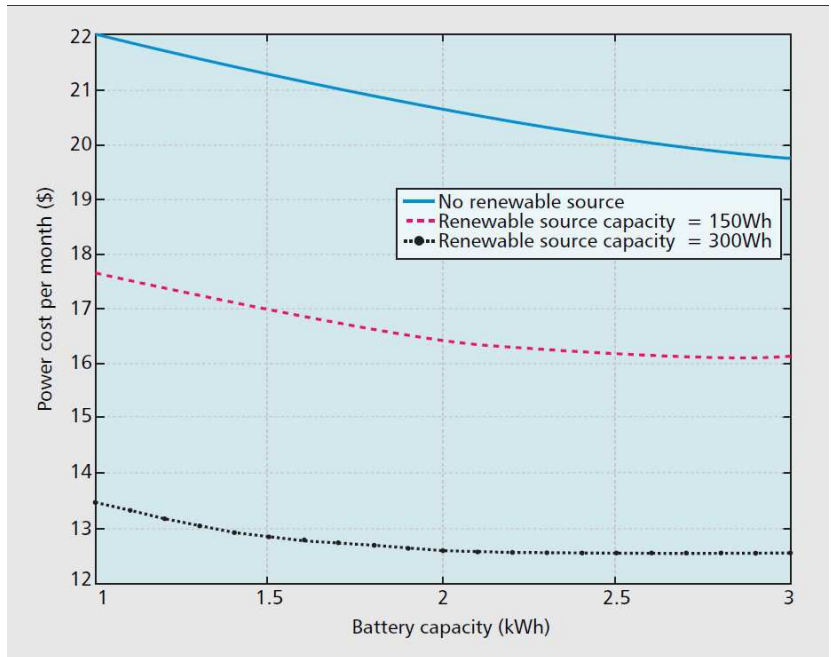}
\caption{Power cost per month under different battery capacity.}
\label{fig:result_cost_battery}
\end{center}
\end{figure}

Next, we study the impact of the battery capacity to the power cost. Fig.~\ref{fig:result_cost_battery} shows the power cost per month under different battery capacity and different renewable source capacity. As the battery capacity increases, the adaptive power management controller can store more power when there is the renewable power generated or when the price of power from electrical grid is cheap. As a result, the power cost per month decreases. However, at a certain capacity, the power cost becomes constant in which all generated renewable power or the power with cheap price from electrical grid can be stored in the battery and sufficient for the future demand. Consequently, increasing the capacity of battery power further will not reduce power cost, while the cost of battery will be higher.

In addition, as expected, when the capacity of renewable power source increases, the power cost buying from electrical grid decreases (Fig.~\ref{fig:result_cost_battery}). However, it is also important to note that the cost of increasing the capacity of renewable power source cannot be ignored. For example, the average price of solar panel with capacity of 150W is 230 dollars. Therefore, the optimal deployment of battery capacity and renewable power source capacity is an important issue and can be studied in the future work.

\begin{figure}[h]
\begin{center}
\includegraphics[width=4.3in]{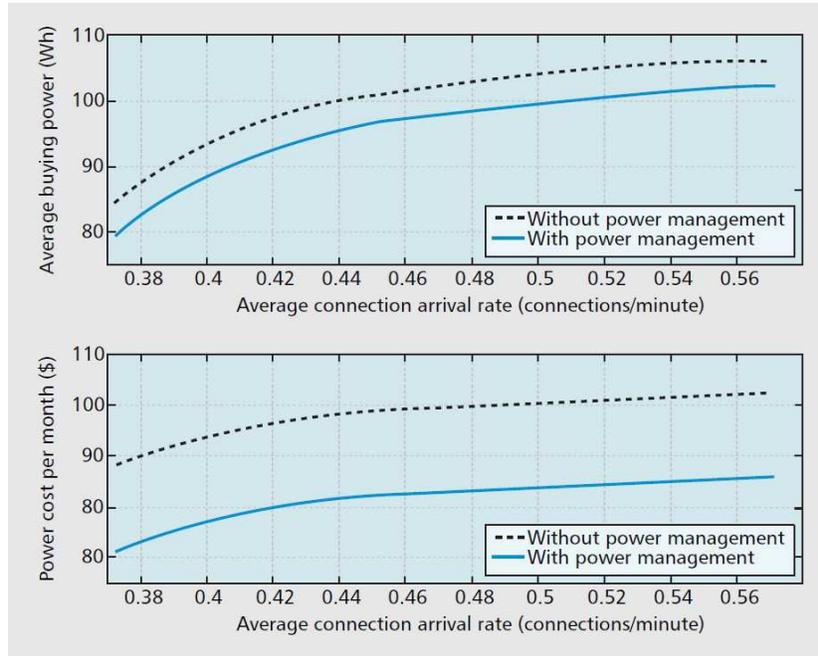}
\caption{Average buying power and battery storage under different connection arrival rates.}
\label{fig:result_power_call}
\end{center}
\end{figure}

Then, we investigate the effect of traffic load of micro base station to the power consumption. Fig.~\ref{fig:result_power_call} shows the average power stored in a battery and power buying from electrical grid under different connection arrival rates. As expected, when the connection arrival rate increases, the micro base station consumes more power. However, with the renewable power source, the consumed power can be supplied partly from electrical grid. Similarly, the power of battery storage increases as the traffic load increases. Due to higher power consumption, adaptive power management controller increases the power storage (e.g., from renewable power source).

Next, the impact of threshold in connection admission control (CAC) to the QoS performance and power cost is studied. With the guard channel CAC~\cite{fang2002}, the threshold is used to reserve the channels for handoff connections, since the mobile users are more sensitive to the dropping of handoff connection than the blocking of new connection. With guard channel CAC, the new connection is accepted if the current number of ongoing connections is less than the threshold, and will be rejected otherwise. As expected, as the threshold becomes larger, more new connections are accepted and can perform data transmission. As a result, the new connection blocking probability decreases. However, handoff connection dropping probability increases, since less channels are reserved. We observe that as the threshold increases, there will be more ongoing connections with micro base station. Therefore, the power consumption increases, and the power cost saving (compared to that without CAC) decreases. This result can be used to optimize the parameter (i.e., threshold) of CAC. For example, if the objective is to minimize the handoff connection dropping probability and maximize the power cost saving subject to the new connection dropping probability to be less than 0.1. Then, the threshold should be set to 20.

From above results, it is clear that the proposed adaptive power management can minimize the cost of power consumption given various uncertainty including renewable power generation, power price, and traffic load of the wireless base station. The optimization formulation will be useful for the design of the resource management of the wireless system in green wireless communications.


\section{Conclusion}
\label{sec:conclusion}

In this article, the adaptive power management for wireless base station incorporating renewable power source in smart grid environment has been proposed. Renewable source offers an alternative power supply which not only saves the power cost, but also reduces the ${\mathrm{CO}}_2$ emission. In addition, with the smart grid, the adaptive power management can be part of a demand side management in which the power consumption can be adapted according to the power supply condition (i.e., power price). In this regard, the optimization problem has been formulated and solved to obtain the optimal decision of adaptive power management. The uncertainty of generated power from renewable source, power price from electrical grid, and power consumption of wireless base station due to varied traffic load has been taken into account. The performance evaluation has been performed and the results clearly show that the optimal decision of adaptive power management can successfully minimize the power cost.

\section{Acknowledgment}
\nonumber

This work was done in the Centre for Multimedia and Network Technology (CeMNet) of the School of Computer Engineering, Nanyang Technological University, Singapore.

\newpage

\begin{figure}[h]
\begin{center}
\begin{eqnarray}
\min_{x_{t,\omega}} & & \sum_{t=1}^T {\mathbb{E}} \left( x_t P_t + s_t L \right) = \sum_{t=1}^T \sum_{\omega \in \Omega } Pr(\omega) \left( x_{t,\omega} P_{t,\omega} + s_{t,\omega} L \right)	\label{eq:objective}	\\
\mbox{subject to}	& & s_{t,\omega} + x_{t,\omega} + R_{t,\omega} = s_{t+1,\omega} + C_{t,\omega} + y_{t,\omega}, \quad t=1,\ldots,T-1, \;\; \omega \in \Omega	\label{eq:const1} \\
					& & s_{t,\omega} \leq B, \quad \quad t = 1, \ldots, T, \;\; \omega \in \Omega		\label{eq:const2} \\
					& & s_{1,\omega} = B_1, \quad s_{T,\omega} = B_T		\label{eq:const3} \\
					& & x_{t,\omega} \geq 0, \quad s_{t,\omega} \geq 0, \quad y_{t,\omega} \geq 0, \quad t=1,\ldots,T , \;\; \omega \in \Omega 	\label{eq:const4}
\end{eqnarray}
\caption{Multi-period stochastic programming model for adaptive power management.}
\label{fig:optimization}
\end{center}
\end{figure}

\begin{figure}[h]
\begin{center}
\includegraphics[width=4.3in]{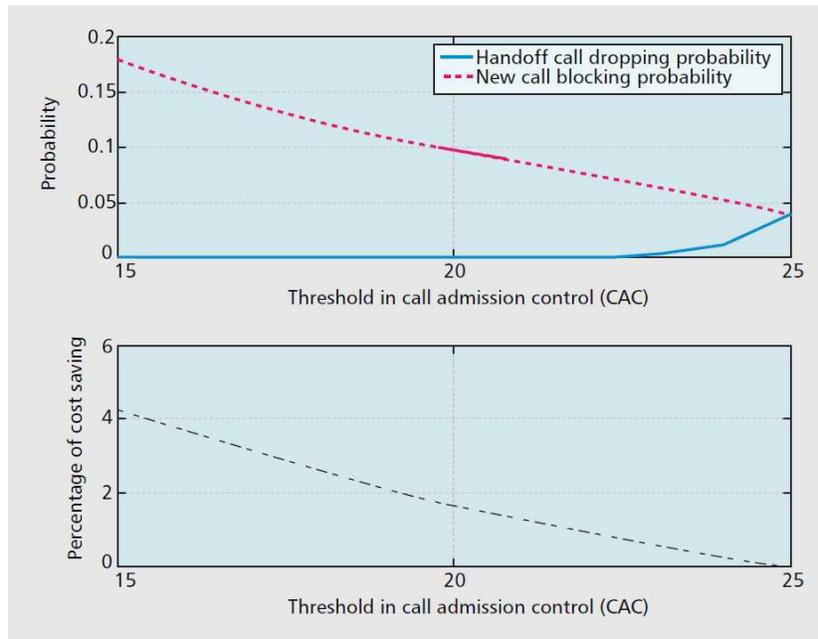}
\caption{Handoff call dropping probability, new call blocking probability, and percentage of cost saving under different threshold in call admission control (CAC).}
\label{fig:result_blockdrop_cac}
\end{center}
\end{figure}

\end{document}